\documentclass[prb,twocolumn,showpacs]{revtex4}
\usepackage[dvips]{graphicx}
\usepackage{latexsym}
\usepackage{color}
\usepackage{amsmath}
\usepackage{bm}
\usepackage{epsfig}
\begin{document}

\title{Topological spin Hall states, charged skyrmions, and superconductivity in two dimensions}

\author{Tarun Grover and T. Senthil}
\affiliation{Department of
Physics, Massachusetts Institute of Technology, Cambridge,
Massachusetts 02139}

\date{\today}
\begin{abstract}
We study the properties of two dimensional topological spin hall insulators which arise through spontaneous breakdown of spin symmetry in systems that are spin rotation invariant. Such a phase breaks spin rotation but not time reversal symmetry and has a vector order parameter. Skyrmion configurations in this vector order parameter are shown to have electric charge that is twice the electron charge. When the spin Hall order is destroyed by condensation of skyrmions superconductivity results. This may happen either through doping or at fixed filling by tuning interactions to close the skyrmion gap. In the latter case the superconductor- spin Hall insulator quantum phase transition can be second order even though the two phases break distinct symmetries.

\end{abstract}

\newcommand{\fig}[2]{\includegraphics[width=#1]{#2}}
\newcommand{\be}{\begin{equation}}
\newcommand{\ee}{\end{equation}}
\newcommand{\wdt}{\widetilde}
\newcommand{\red}{\textcolor{red}}
\maketitle

Several recent theoretical\cite{kanemele1, kanemele2, bern06} and experimental\cite{konig07} papers  have discussed a phenomenon known as the quantum spin Hall effect in two dimensional insulators. In such an insulator an applied electric field leads to a quantized spin current in the transverse direction. Clearly the effect cannot occur if $SU(2)$ spin rotation symmetry is preserved in the low energy theory of the insulator. Initial discussions of the spin Hall effect focused on situations where the microscopic Hamiltonian has spin-orbit interactions which are such that only the $S^z$ component of the spin is conserved. Alternately one can contemplate phases of matter where full $SU(2)$ spin rotation symmetry is present at the microscopic level but is spontaneously broken in the ground state in a manner that  enables a spin Hall effect. Such phases were discussed recently in Ref. \onlinecite{raghu}.

In this paper we study various aspects of the second kind of spin Hall insulator where the spin symmetry is spontaneously broken. The broken spin symmetry is characterized by a vector order parameter which we will denote
$\vec N$. Consequently in two space dimensions skyrmion topological defects are allowed. We show that the quantized spin Hall effect leads (with some restrictions discussed below) to a quantized electric charge $2e$ on the skyrmion. Here $e$ is the electron charge. This has some remarkable consequences.
First so long as the skyrmion energy is much smaller than the energy to excite individual pairs of electrons, the skyrmion number will be conserved due to charge conservation. Thus space-time hedgehog configurations of $\vec N$ are forbidden\cite{footnote1}. Destruction of the spin Hall order by condensation of skyrmions then very simply leads to a gapped $s$-wave superconductor. This may be done either by doping into the spin Hall insulator or by closing the skyrmion gap at fixed density by tuning interactions. In the latter case
the resulting superconductor - spin Hall insulator phase transition can be second order despite their rather different broken symmetries. This provides a new example of a Landau-forbidden deconfined quantum critical point \cite{deccp}.

These results are best illustrated in the context of models of
of spinful electrons $c_{\bm{r}\sigma}$ hopping on a honeycomb lattice at half-filling:
\begin{equation}
H = \sum_{<\bm{r}\bm{r'}>} -t\, (c_{\bm{r}}^{\dagger} c_{\bm{r'}} + h.c.) + H_{int},
\end{equation}
where $\bm{r}, \bm{r'}$ are nearest neighbor sites. $H_{int}$ contains various short-ranged interactions that
preserve spin rotation and time reversal symmetries. Specific interactions that could stabilize the
spin Hall insulating phase are discussed in Ref. \onlinecite{raghu}. In the absence of $H_{int}$,
the band-structure consists of two distinct Fermi \textit{points} which can be chosen as $ \bm{K} = (\pm 4 \pi/{3\sqrt{3}}, 0) $ where the lattice constant is chosen as unity.
Expanding the microscopic electron annihilation operator $c_{\bm{r}}$ about these nodes in
terms of continuum field operators $c_{i\alpha\sigma}$  the low energy Dirac Hamiltonian is obtained as
\begin{equation}
H = -\frac{3t}{2} \int d^2 k \, c^{\dagger}(-\tau^{x} \mu^{z} k_x + \tau^{y} k_y)c
\end{equation}
Here $i = L/R$ is a sublattice index, $\alpha =
 +/-$ the node index, and $\sigma$ the spin.  The corresponding Pauli
 matrices are denoted  $\bm{\tau}, \bm{\mu}$ and $\bm{\sigma}$ respectively.
Further the transformation $c_{+} = \tau^{y} \psi_{+}$ enables writing the
the Hamiltonian in a manifestly symmetric form
\begin{equation}
H  =  -\frac{3t}{2}\, \int d^2 k \,{\psi}^{\dagger} (k_x {\tau}^{x} + k_y {\tau}^{y}) \psi
\end{equation}
The corresponding action (in real space) is
\begin{eqnarray}
S & = & \int d^3x \overline\psi \left(-i \gamma^{\mu} \partial_{\mu}  \right ) \psi
\end{eqnarray}
Here $\mu = \tau,x,y$ with $\tau$ being imaginary time,
$\overline\psi = -i\psi^{\dagger} \tau^z $ and ${\gamma^0, \gamma^x, \gamma^y} = {\tau^z,  \tau^y,  - \tau^x}$. We have rescaled space-time to set the Dirac velocity to $1$.

Consider a phase where the interactions $H_{int}$ leads to a non-zero expectation value for the operator $\overline{\psi} \bm{\sigma} \psi$, say along the $\hat{z}$ direction in the spin-space:
 \begin{equation}
 <\overline{\psi} \bm{\sigma} \psi> \equiv \vec N~~= N^z \hat{z}
 \end{equation}
 Within a mean field description this leads to a mass for the low energy Dirac fermions which is opposite for $\uparrow$ and $\downarrow$ spin:
 \begin{equation}
 S_{MF} = \int d^3x \overline{\psi}\left(-i \gamma^{\mu} \partial_{\mu}  + i m \sigma^z \right ) \psi
 \end{equation}
 where the mass $m = \lambda N^z$ with $\lambda$ determined by the interactions that lead to this order.
 At the mean field level this is identical to the model discussed in Ref. \onlinecite{kanemele1} for the quantum spin hall effect. It breaks spin rotation but not time reversal symmetry. In a sample with a boundary, an electric potential difference $V$ applied between the right and left edge leads to a pair of propagating edge states with opposite spin \cite{kanemele1, kanemele2, bern06}. Each spin species carries a non-vanishing charge current of magnitude $\frac{e^2}{h} V$ transverse to the electric field in opposite directions. This corresponds to a spin-hall conductivity $\sigma^s_{xy}= \frac{2e^2}{h} \left(\frac{\hbar/2}{e}\right) = e/2\pi$.

It is useful to characterize the spin hall effect in terms of the response to two different external gauge fields $A^c$ and $A^s$ which couple to spin and charge currents respectively. Consider the mean field action in the presence of these gauge fields
\be
S_{MF}[A^c, A^s] = \int d^3 x \, \overline\psi (\gamma_{\mu}(-i\partial_{\mu} + {A^c}_{\mu} + \frac{\sigma^z}{2} {A^s}_{\mu})) \psi + im  \overline\psi \sigma^z \psi \label{eq:chern1}
\ee
in the units $e = \hbar = 1$. When the fermions are integrated out, a non-vanishing transverse spin-hall conductivity of magnitude $1/2\pi$ implies that the low energy effective action for the gauge fields is given by
\be
S_{eff} = \frac{i}{2\pi} \int d^3 x \, \epsilon^{\mu \nu \lambda} A^c_{\mu} \partial_{\nu} A^s_{\lambda} \label{eq:s_eff}
\ee

Consider now the effect of fluctuations. These may be usefully discussed by considering the Dirac action in the presence of a fluctuating  unit vector field $\hat{N}$ describing the orientation of the spin Hall order parameter:
\be
S = \int d^3 x \,  \overline\psi (-i \gamma_{\mu}\partial_{\mu}    + im \vec \sigma .\hat N) \psi \label{eq:S_nonfluct}
\ee
We may now integrate out the fermions. In the limit of large mass $m$ the result is a non-linear sigma model:
\be
S = \int d^3x \frac{1}{g} \left(\partial_{\mu} \hat{N} \right)^2 + ....
\ee
with `stiffness' $\frac{1}{g} \sim |m|$. The ellipses represent higher order terms in the $1/m$ expansion.
Clearly in the ordered phase there will be two gapless  linear dispersing Goldstone modes associated with the broken spin symmetry.

The other class of excitations associated with the vector order parameter $\hat{N}$ are skyrmion configurations.
The quantum numbers of a skyrmion whose size is much bigger than the length scale $1/m$ can be conveniently discussed within an adiabatic approach where such a skyrmion is slowly built up from the ground state. We now show that such a `fat' skyrmion has electric charge $2$.

To see this most simply, lets consider the Dirac action Eqn. \ref{eq:S_nonfluct} in the presence of a static background configuration of the $\hat{N}$ field that corresponds to a single skyrmion.
An example is the configuration
\begin{equation}
\hat{N}(r, \theta) = \left(sin(\alpha(r)) cos(\theta), sin(\alpha(r)) sin(\theta), cos(\alpha(r))\right)
\label{skyrme}
\end{equation}
with the boundary conditions $\alpha(r=0) = 0$ and $\alpha(r \rightarrow \infty) = \pi$. Here $(r, \theta)$ are polar coordinates for two dimensional space. This field configuration corresponds to a skyrmion with Pontryagin index one.
Rotate the $\hat{N}$ vector field to the $\hat{z}$ direction by a unitary transformation $U$ at all points in space: $U^{\dagger} \left(\vec\sigma.\hat{N}\right) U = \sigma^z$. Further, one may define $\psi = U \psi'$ so that Eqn. \ref{eq:S_nonfluct} becomes
\be
S = \int d^3 x \,  \overline{{\psi}'} \left(\gamma_{\mu} \left(-i\partial_{\mu}  + B_\mu \right)  + im \sigma^z \right) {\psi}'
\ee

where $B_\mu = -i U^{\dagger} \partial_\mu U$.  One readily finds that in the far field limit $\vec{B} (r \rightarrow \infty, \theta) \rightarrow \sigma^z \frac{\hat{e}_\theta}{r}$ where $\hat{e}_\theta$ is the unit vector along $\theta$ direction. Therefore a skyrmion with unit Pontryagin index induces a spin gauge field $\vec{A^s} = 2 \hat{e}_\theta/r$.  The total flux of this gauge field is $4 \pi$ which remains invariant under smooth deformations of the $\hat{N}$ field (which does not change the skyrmion number).

Let us gradually build in a skyrmion configuration by starting from the ground state. By the above argument, this threads in a spin gauge flux of $4\pi$ for $A^s$. Due to the spin-hall effect of the mean field state described above, this would result in a flow of electric current of magnitude $j^c =\frac{1}{2\pi}\frac{ \partial{A^s}}{\partial t}$ in the radial direction. The total charge transferred from the center of the sample to the boundary during the process equals $Q^c = \int_C j^c\, r d\theta$ where the contour $C$ is a circle located near infinity. This then leads to a charge $Q^c = 2 e$ associated with the skyrmion.  \textit{Thus a skyrmion with Pontryagin index one carries an electrical charge}  {$2e$}.

This physical argument is confirmed by a more sophisticated analysis which carefully integrates out the Dirac fermion fields in the presence of an external charge gauge field. Specifically we consider
\be
S[A^c] = \int d^3 x \,  \overline\psi (\gamma_{\mu}\left(-i \partial_{\mu} + A^c_{\mu}\right)  \psi + im \overline\psi \vec \sigma .\vec N) \psi
\ee
Integrating out the fermions and using the large mass expansion yields\cite{abawieg99,abawieg01}
\be
\label{sigmaAc}
S = \int d^3x \frac{1}{g}(\partial_\mu N)^2 + 2iA^c_\mu J^T_{\mu}
\ee
where $J^T_{\mu} = \frac{1}{8\pi} \epsilon^{\mu \nu \lambda} \vec{N}.\partial_\nu\vec{N} \times \partial_\lambda\vec{N}$ is the topological current whose time component equals the skyrmion density associated with the order parameter field \cite{footnote2}. It follows that the skyrmion carries charge which is given by

\be
Q^c_{skyrmion}   =  2e \int dx dy  \frac{1}{8\pi} \epsilon^{0 \nu \lambda} \vec{N}.\partial_\nu\vec{N} \times \partial_\lambda\vec{N}\\
\ee
Thus one again reaches the conclusion that a skyrmion with Pontryagin index $1$ carries an electric charge two.

It is important to emphasize that both the adiabatic argument or the field theoretic derivation above is really valid only for fat skyrmions with size much bigger than $1/m$. It is precisely such fat skyrmions that are important for our considerations below.

The electric charge of the skyrmions has profound implications for  quantum phase transitions out of the spin hall insulator. First if the energy of a fat skyrmion is much smaller than the energy of individual pairs of electrons, then the skyrmion number is conserved. Thus hedgehog configurations of the $\hat{N}$ vector in spacetime (which correspond to events that change the Pontryagin index) are prohibited. Further in the absence of any topological terms in the effective action for the $\hat{N}$ field, the skyrmions will be bosons \cite{footnote2}. Consider therefore the result of  condensing the skyrmions by tuning the interactions in $H_{int}$ while keeping the electron density fixed. Due to the charge on the skyrmions the result will be a superconductor! This superconductor will have a single electron gap, and flux quantization in units of $\frac{hc}{2e}$. Indeed it is an ordinary $s$-wave superconductor.

Remarkably the quantum phase transition between the spin hall insulator and the superconductor can be second order.
This is despite the rather different broken symmetries in the two phases. This is because the skyrmion condensation transition can simply be understood as the phase transition of an $O(3)$ vector model in the absence of hedgehogs.
This is conveniently described in the $CP^1$ formulation of the $\hat{N}$ field by writing $\hat{N} = z^{\dagger} \vec \sigma z$ field with $z$ a two component complex spinon field:

\be
\label{CP1}
S_{eff} = \frac{1}{g} \int d^3 x \,  |(-i\partial_\mu - a_\mu)z|^2
\ee
Here $z = (z_1, z_2)$  are coupled to an {\em internal} gauge field $a_{\mu} = -iz^{\dagger}\partial_{\mu} z$. The skyrmion number $Q$ is simply given by
\be
Q = \frac{1}{2\pi} \int d^2x \epsilon_{0\mu\nu} \partial_{\mu} a_{\nu}
\ee
 Since the skyrmion number is proportional to the flux of the $CP^1$ gauge field, $a_{\mu}$ of Eqn. \ref{CP1} must be regarded as non-compact, and the phase transition is described by that in the non-compact $CP^1$ model\cite{mv} ($NCCP^1$).

Exactly the same field theory describes the deconfined quantum critical point separating Neel and
Valence Bond Solid (VBS) phases of {\em insulating} square lattice quantum antiferromagnets \cite{deccp}. However there is an important difference: for the Neel-VBS transition quadrupled hedgehogs are actually allowed but have been argued to be irrelevant at the critical fixed point of the $NCCP^1$ model. They are however relevant at the paramagnetic `free photon' fixed point. This leads to the appearance of two diverging length scales at the Neel-VBS transition. In contrast in the present problem the skyrmion number conservation is an exact symmetry at energy scales below the single electron gap which is large and finite in both phases and at the transition. Thus a suitable theory that captures the low energy physics of both phases and their transition should simply ignore all hedgehogs. Consequently there is only a single divergent length/energy scale at the deconfined critical point separating the spin hall insulator from the superconductor.

The competition between the spin Hall order and the superconducting order can also be fruitfully described in a different manner which sheds further light on the role played by the quantum nature of the topological defects near the phase transition. We combine the three component spin Hall order parameter $\vec{N}$ with the two component superconducting order parameter $\psi_{SC}$ into a five component `superspin' vector $\phi_a = (N^x, N^y, N^z, Re \psi_{SC}, Im \psi_{SC})$, and impose a unit length constraint on $\phi_a$, {\em i.e} $\sum_{a = 1}^5 \phi_a \phi_a = 1$.
What is the structure of a
`sigma model' that describes the dynamics of $\phi_a$?   This question may again be answered by coupling the underlying Dirac theory to this superspin vector and integrating out the fermions. Specifically we consider the Dirac
action
\begin{equation}
S = \int d^3x \overline\psi (-i \gamma_{\mu}\partial_{\mu}    + im \vec \sigma .\vec N) \psi + im (\psi_{SC}^* \, \,{\psi}^T \tau^y \sigma^y \mu^y \psi
+ c.c)
\end{equation}
where the last term indicates the coupling to the superconducting order parameter $\psi_{SC}^* \propto \left\langle {\psi}^T \tau^y \sigma^y \mu^y \psi \right\rangle$.
It is potent to make a unitary transformation on the fermions in the $\psi$ basis as  $\wdt{\psi} = e^{i\tau^x \pi/4} \psi$ and expand $\wdt{\psi}$ in its Majorana components $\wdt{\psi} = \eta_1 + i\eta_2$. In the $\eta$ basis the kinetic part of the Dirac action has manifest $SO(8)$ symmetry under rotations of all $8$ components ($2$ spin, $2$ node and $2$ Majorona indices). The spin-hall order parameter is given as
$\left\lbrace \eta^{T} \tau^y \sigma^x \eta, \eta^{T} \tau^y \sigma^y \rho^y \eta, \eta^{T} \tau^y \sigma^z \eta \right\rbrace $ while the s-wave SC has the form $\left\lbrace \eta^{T} \tau^y \sigma^y \mu^y \rho^x \eta, \eta^{T} \tau^y \sigma^y \mu^y \rho^z \eta \right\rbrace$. Here $\bm{\rho}$ are Pauli matrices acting on the Majorana index. These five components together transform as a vector under an $SO(5)$ subgroup of $SO(8)$.  This algebraic structure largely determines important features of the effective sigma model that describes the fluctuations of the $\phi_a$ when the fermions are integrated out. Within the large mass expansion the techniques of Ref. \onlinecite{abawieg99}
lead to an $SO(5)$ action with a Wess-Zumino-Witten(WZW) term at level-$1$.
\begin{equation}
S  =  \int d^3x \frac{1}{2G} (\partial_i \phi_a)^2 + 2\pi i\Gamma\left[ \vec \phi \right]
\end{equation}

The WZW term $\Gamma$ is defined as follows. The field $\phi_a$ defines a map from spacetime $S^3$ to $S^4$. $\Gamma$ is the ratio of the volume in $S^4$ traced out by $\phi_a$ to the total volume of $S^4$.
Specifically let $\hat{\phi}(x,u)$ be any smooth extension of $\hat{\phi}(x)$ such that $\hat{\phi}(x,0)
= (1,0,0,0,0)$ and $\hat{\phi}(x,1) = \hat{\phi}(x)$. Then
\begin{equation}
\Gamma  =  \frac{ \epsilon_{abcde}}{\textrm{Area}(S^4)}
\int_{0}^{1} d\,u \int d^3 \, x \phi_a \partial_x\phi_b \partial_y\phi_c \partial_\tau \phi_d \partial_u \phi_e
\end{equation}
The presence of this `topological term' crucially alters the physics from naive Landau-like descriptions of the competition between the two different orders. Parenthetically we note that the action above must be supplemented with some anisotropy between the $\vec N$ and $\psi_{SC}$ which breaks the $SO(5)$ symmetry down to $SO(3) \times U(1)$. The $SO(3)$  corresponds to
spin rotations and the $U(1)$ to charge conservation symmetry. Exactly the same sigma model with the WZW term also describes\cite{th,sf} the Neel-VBS transition of insulating quantum magnets in agreement with our previous identification.

The physical implications of the WZW term are brought out by the following instructive calculation. Consider a $hc/2e$ vortex defect in the superconducting state. In such a configuration the $5$-component unit vector will point along the $4,5$ directions far from the vortex core but will point along the $1,2,3$ direction in the core. Specifically consider a configuration (we use polar coordinates $(r,\theta)$ for two dimensional space and $\tau$ for time)
\begin{eqnarray}
\hat {\phi} (r, \theta, \tau, u) & = & (sin(\alpha(r,u)) \hat{N}(\tau,u), \nonumber \\
                                 & &  cos(\alpha(r,u)) cos(\theta), cos(\alpha(r,u)) sin(\theta))
\end{eqnarray}

with $\hat{N}^{2} =1$. Choosing $\alpha(r = 0,u) = \pi/2, \alpha(r = \infty, u) = 0  \,\,\, \forall u \ne 0$ and $\alpha(r,0) = \pi/2$, this configuration indeed describes a static superconducting vortex where
$\hat{\phi}$ points along the $\vec N$ direction in the core. We have allowed the unit vector $\hat{N}$  in the core to have time dependence. The integrals defining $\Gamma$ are readily evaluated for this configuration and lead to the result
\begin{equation}
\Gamma = \frac{1}{4\pi}\int d\tau du \hat{N}. \partial_{\tau} \hat{N} \times \partial_u \hat{N}
\end{equation}
This is {\em precisely} the quantum Berry phase for a spin-$1/2$ object. Thus we see that {\em close to the transition} the superconducting vortices behave as spinons. Condensing the vortices destroys the superconducting order but at the same time leads to condensation of the spin Hall order. This is the mechanism for the Landau-forbidden transition. The arguments of Ref. \onlinecite{z4vrtx} now establish the equivalence of the sigma model to the $NCCP^1$ field theory and the superconducting vortices are directly identified with the $CP^1$ spinons $z$.

Finally we consider the result of doping the spin Hall insulator focusing again on the situation where the energy of a fat skyrmion is much smaller than the energy gap of individual electrons. Then the doping will
be accommodated by the introduction of a finite density of skyrmions into the system. At very low doping a skyrmion crystal will presumably be stabilized but with increasing doping a translation invariant
skyrmion condensate is expected. This will be a superconducting phase with no broken spin rotation symmetry.

The route to superconductivity discussed in this paper is rather different from two common theoretical mechanisms
 - phonon or other boson mediated pairing of electrons, or resonating valence bond pairing originating from superexchange in a Mott insulator. Rather it involves condensation of charged solitons of a different broken symmetry, namely the spin Hall order. Similar phenomena have previously been considered in field theoretic contexts\cite{wiegmann}, and is known as topological superconductivity. Superconductivity arising from the quantum spin Hall state thus provides a nice realization of a certain kind of topological mechanism for superconductivity.

This work was supported by NSF Grant DMR-0705255.


\begin{references}
\bibitem{footnote2} Note that a Hopf term will also be generated in the large mass expansion; however its coefficient is $\theta = \pi N_f$ where $N_f$ is the number of species of fermions \cite{abawieg99} and $N_f = 2$ here. It can therefore be dropped for the subsequent discussion. In particular the skyrmion statistics will be bosonic.
\end{references}
\end{document}